# Transaction Costs: Economies of Scale, Optimum, Equilibrium and Efficiency

A game theory-based model of transaction costs

László Kállay, Tibor Takács, László Trautmann[1]

*Project no. NKFIH-869-10/2019 has been implemented with support provided from the National Research, Development and Innovation Fund of Hungary, financed under the Tématerületi Kiválósági Program funding scheme.*

## Abstract

The aim of this article is to propose a core game theory model of transaction costs wherein it is indicated how direct costs determine the probability of loss and subsequent transaction costs. The existence of optimum is proven, and the way in which exposure influences the location of the optimum is demonstrated. The decisions are described as a two-player game and it is discussed how the transaction cost sharing rule determines whether the optimum point of transaction costs is the same as the equilibrium of the game. A game modelling dispute between actors regarding changing the share of transaction costs to be paid by each party is also presented. Requirements of efficient transaction cost sharing rules are defined, and it is posited that a solution exists which is not unique. Policy conclusions are also devised based on principles of design of institutions to influence the nature of transaction costs.

## 1. Introduction

People around the world make billions of transactions on an average day most frequently, but not limited to payments. We must execute transactions to acquire commodities, buy or sell our cars and houses, getting a job and being paid for our work. Executing even the smallest transactions bears risk therefore managing risk is a daily routine for us. But how efficiently do we do all this? Can the cost of executing transaction be lower? What does it depend on if we can reduce transaction costs? Are only our decisions responsible for the price we must pay, or do the rules and norms we follow also count?

The cost of transactions itself is an important question, it has an impact on living standards, but transaction costs also influence how efficiently we can utilize our resources. Understanding what determines the price we pay for the transaction we make is of crucial importance.

The goal of this paper is to propose a game theory model of transaction costs. In the theoretical framework of the model among others the following will be proven.

(a) Transaction cost has an optimum determined by the connection between the direct costs and the level of losses caused by uncertainty. (b) The optimum point and value are a function of exposure. (c) Executing the transactions can be modelled with a two-player game. (d) Whether the equilibrium of the

---

game is the optimum point of the game depends on the rules or institutions players must follow. (e) The transaction cost sharing rule has a regulatory function.

Executing a transaction is a risk management problem implying the possibility and need of optimization. The efforts taken by the actors to reduce loss have an impact on the probability of loss and not on the absolute amount. The decisions made by the actors executing a transaction are influenced by the rules the actors must follow.

By challenging assumptions of the neoclassical paradigm, the importance of transaction costs is acknowledged and further discussed with reference to recent literature on institutional economics. However, it would seem only partial attempts at formally modelling decisions determining levels of transaction costs have been made.

In this paper, a core model of transaction costs is proposed indicating how direct costs determine the probability of loss and subsequent transaction costs. The existence of optimum is proven as well as how exposure influences where the optimum point of decisions is located. Decision making is described as a two-player game prior to discussion of how the transaction cost sharing rule determines whether the optimum value of transaction costs is the same as the equilibrium value of the game. A game modelling disputes of actors in terms of negotiating the share of cost to be paid by respective players is also presented.

The model does not discuss resource or property rights allocation, but rather focuses on how transaction costs are determined. Finally, conclusions are drawn as to how efficient institutions should and could be designed.

Organization of the paper commences with an overview of the relevant literature. The transaction costs model and the model of changing the share of costs through disputes are presented in the subsequent section. Section 4 is devoted to policy conclusions, while general conclusions are presented in Section 5.

## 2. Literature overview

For the purpose of this paper the notions of both the old and the new institutional economics and transaction cost economics are overviewed having relevance for our model development.

In terms of economic theory, a shift of emphasis from the exchange of commodities to that of a concept of transaction began at the beginning of the 20th century. The latter approach formed part of the neo-classical revolution whereby the emphasis of economy theory development shifted from value of labor theory towards that of utility theory.

The notion of transaction as a component of the institutional approach was introduced by Commons (1924), where the basic institutional framework of the market was examined assuming individuals pursue self-creation as opposed to utility which the market can support if well-designed. In contrast to mere exchange, transaction forms in essence the mutual agreement of partners to take efforts willingly to exchange goods or services. Physical force and coercion may also be used to execute transactions. Commons argued that such a tendency was one of the reasons why political and economic turbulence existed in the USA in the late 19th and early 20th centuries. Legal decisions made by relevant institutions in that period influenced overall movement in the composition of the economy and the exchange mechanism towards that of a 'free will' basis.

An accepted starting point of discussion of transaction costs is encapsulated in Coase's (1937) response to the question as to why it is profitable to establish a firm, whereby it is posited that use of the price mechanism entails costs. Coase (1937) also discusses the problem of minimizing contract costs and the possibility of selecting cheaper means of contracting. In a later paper (Coase 1960), it is suggested that bargaining will result in a Pareto efficient outcome of allocation of property rights provided transaction costs are sufficiently low. Coase (1960) also noted the existence of transaction costs by pointing out that governments may not necessarily contribute to mitigation of their effects. This element of theory underlined the importance of transaction costs and raised the question as to how transaction costs may affect property rights allocation in resource terms if not seen to be sufficiently low.

Economic theory is quite clear in this regard in that there is no resource allocation process without entailing a transaction cost. Correspondingly, control of transaction costs is a perennial subject of economic political discussion.

Latter developments in economic theory relied heavily on Coase's transaction theory. This is embedded particularly in contract theory (Hart and Holstrom 1986) and in organizational theory (Williamson 1979). Other scholars have also discussed the role of transaction costs and factors which may determine their role and extent.

More recently, the concept of transaction has been extended beyond that of a narrow frame of reference of market-based exchange. This has been interpreted for use in several different types of behavioural approaches with the shift partly based on Williamson's (1979) work, whereby dimensions for characterizing transactions are defined as being of frequency, specificity and uncertainty. Opportunistic behaviour and bounded rationality also form central concepts in the study of transaction costs to thus explain why contracts are necessarily incomplete. Williamson also point out that reduction in monitoring generally gives more space for opportunism inducing thinking about optimizing transaction costs.

The role of institutions in minimizing transaction costs has often been discussed by authors adhering to the school of New Institutional Economics. Besides several other factors, North (1994) suggests that institutions shape economic performance and that information is not only costly but also incomplete, resulting in costly imperfect enforcement.

Summarizing literature overviewed earlier, transaction costs can be divided into three broad categories as follows (Roberts and Milgrom 1992):

- Search and information costs which may include for example whether the required good or service is available on the market and at the lowest price.
- Bargaining costs required to arrive at an acceptable agreement with the other party involved in a transaction in order to construct a mutually amenable contract.
- Policing and enforcement costs for ensuring alternate parties act in compliance with the terms of contracts and taking appropriate legal action if this does not transpire in practice.

Anderlini and Felli (2006) challenged the robustness of the Coase theorem through proposing a model with two parties which can generate an economic surplus by means of execution of a transaction. However before negotiations begin on dividing the surplus, each party should incur transaction costs. This may entail a 'hold-up problem' if there is a mismatch between the bargaining power of respective parties and between the magnitude of transaction costs. If one party possesses large transaction costs but in future negotiations it is only able to seize a relatively small fraction of the surplus through relatively low bargaining power, then it will not incur transaction costs and hence the total surplus will be forgone.

Thus, the importance of transaction costs is underlined indicating that a certain type may be an obstacle to efficient resource allocation.

Further cooperative game theory models of the Coase theorem either assume that there is no transaction cost (Gonzalez et al. 2019) or assume it is a fixed amount (Robson 2019) and do not address the problem what factors determine the level of transaction costs.

# 3. Model development

## 3.1. Definitions

The notion of transaction has several slightly different interpretations in terms of 20[th] century discourse on economic theory. For the purpose of this paper, transaction is defined as the act of exchange and is seen as a precondition of division of labour and of specialization. As such, it may be considered as an integral functional component of modern economies.

Moreover, the total cost of transactions is considered as a prime explanatory factor for the set of feasible acts of exchange wherefrom viable transformation may ensue in the form of production activities.

Crucially, *Actors* are defined as individuals or organizations interested in executing transactions. In this paper a two-actor model is presented and discussed. During execution, the good or service subject to the transaction is exposed to the possibility of loss which may result from factors such as physical damage, fraud, administrative or technical errors or additional operational costs. The amount of loss may exceed the value of the transaction if for example, either goods incur physical damage during transportation or alternately a fraudulent transfer of funds is debited from an actor's account resulting in additional problem resolution efforts.

On this basis, the value at stake when executing a transaction, or the possible maximum loss is defined as *exposure* and is denoted by $e > 0$. Efforts of actors to reduce the risk of the transaction may affect the probability but not the absolute value of loss. Actors may possess a finite number of choices in terms of efforts, actions and measures which can be taken to decrease the probability of loss denoted by $ch_i^1, ch_j^2 \ i = 1, \dots, n; \ j = 1, \dots, m$. An example of possible choices faced by actors may be the elements of authentication of payment transactions[2] that is (1) self-contained and undivulged knowledge, (2) sole possession and (3) self-contained inherence in the form of actual existence of the user. Requirement of additional elements to authenticate payments increases levels of effort expended by actors while the probability of unauthenticated financial transfers is decreased. To enhance the security of payment transactions, more elements of the same category may also be required such as the use of single use passwords as authentication codes to result in further increases in the cost of a transaction and decreases in the probability of loss.

Actors may make decisions on which action they take either simultaneously or sequentially. Not all the combinations or pairs of efforts are possible and some of the combinations may result in cancellation of the transaction. The set of possible combinations is denoted by $Pch$, where $\left(ch_i^1, ch_j^2\right) \in Pch$ is defined as a possible pair of choices. Costs of efforts are transaction specific since they are not recoverable and cannot be transformed into resources for any other purpose. These are often referred to as sunk costs and are considered here as the *direct costs* of a transaction.[3] The direct costs depend on the choices made by actors and are denoted by $z_1(ch_i^1), z_2(ch_j^2) \geq 0$.

The *probability of loss* (loss ratio) $0 \leq Pl(ch_i^1, ch_j^2) \leq 1$ also depends on choices made by actors provided $(ch_i^1, ch_j^2) \in Pch$ otherwise the transaction cannot be executed. The loss ratio and exposure level determine how the decisions of actors effect the expected loss of the transaction as part of the total cost. The expected loss is therefore $L = Pl(ch_i^1, ch_j^2)e$.

Furthermore, expected loss can be defined as quantification of the consequences of uncertainty which is viewed by Williamson (1979) as one of three dimensions required for characterizing transactions. In this model, the direct costs are (negative) utilities, and the expected loss is correspondingly an expected utility.[4] So, they can be measured by the same dimension, we can add them. Since a given extent of loss is inevitable, the *total cost of executing the transaction* (hereafter referred to as transaction cost) is the sum of transaction specific or direct costs and the loss, incurred while executing the transaction $Tc(e, ch_i^1, ch_j^2) = z_1(ch_i^1) + z_2(ch_j^2) + Pl(ch_i^1, ch_j^2)e$ within domains of possible pairs of choices $(ch_i^1, ch_j^2) \in Pch$.

*Transaction type* ($Ttype$) is defined as the set of transactions where the set of choices $(ch_i^1, ch_j^2)$, the direct cost of choices $z_1(ch_i^1), z_2(ch_j^2)$, and the pairs of possible choices ($Pch$) are exactly the same and choices of actors result in the same probability of loss ratio $Pl(ch_i^1, ch_j^2)$. In other words, transactions belonging to the same transaction type may defer only in the extent of exposure. An example may be in paying for fast moving consumer goods at shops with a bank card in one concrete country, or transferring money using internet banking or in payment for goods ordered online. With reference to Williamson's (1979) categorization of frequency as one of the three critical dimensions of transactions, the model used here defines frequency in a wider sense given there can be many individuals or organisations involved as actors in a given form of transaction while only the extent of exposure may vary.

## 3.2.   Existence of optimum transaction cost

On this basis, the question of how the transaction cost can be minimized is investigated.

One possible combination of choices determines direct costs of the choices of actors, $z_1(ch_i^1)$, $z_2(ch_j^2)$, and the loss ratio and resultant transaction cost wherein exposure is given in a single transaction.  The same direct cost combinations may be the result of different pairs of choices causing different levels of loss ratio. Some of the possible combinations may not be optimal at any level of exposure and will be eliminated algorithmically as subsequently outlined. Firstly, all direct cost combinations resulting in higher loss ratios are eliminated. Thus, the probability of loss as a function of direct costs can be examined.

---

[3] Transaction specific costs in the model differ from transaction specific investment as used in the context of asset specificity as defined by Williamson (1975, 1979, 1985), because these are only direct costs related to the execution of the transaction. They also differ from transaction specific governance structures since they only refer to the cost of operating them in individual occasions.

[4] Actors preferences in terms of cost and loss might be different, some may prefer avoiding losses, while others value saving direct costs higher. For the sake of simplicity, we assume that actors have risk neutral preferences.

The probability of loss or loss ratio function is given as $Plf(z_1, z_2): D \to [0; 1]$, where $D \subset R^2_{\geq 0}$, $z_1$ and $z_2$ are costs assigned to possible pairs of choices of actors. The number of possible decision pairs or combination of choices is finite. The loss ratio and exposure levels determine how the decisions of actors affect the expected transaction loss as a proportion of the total cost, given by:

$$Tcf(e, z_1, z_2) = z_1 + z_2 + Plf(z_1, z_2)e.$$

The quantity $(1 - Plf(z_1, z_2))e$ measures the yield of the decision combination. Only relevant pairs are considered. Specifically, these are pairs which may result in optimum or minimum transaction costs at a given exposure level. In the range of the loss ratio, the values of the function thus lie on a surface, which strictly decreases by both dimensions and is convexity as illustrated in figure 1.

A greater amount of expenditure yields a greater amount of loss reduction but the yield decreases in direct proportion to the principle of diminishing returns. Relevant pairs can be selected in such a way that those which are non-effective are omitted. This can be performed by use of the standard data envelopment analysis (DEA) method (Seiford and Thrall 1990; Seiford and Zhu 1999) which may determine the most efficient decision-making units (DMUs) relative to all other units. It is further assumed that each unit possesses the same types of inputs and outputs. In this present transaction cost model, possible pairs of decisions perform the role of DMUs, where the coordinates $z_1$ and $z_2$ act as inputs, and the yield of decision $e(1 - Plf(z_1, z_2))$ can be considered as outputs. The set of relevant points can be determined by solving a linear programming problem for each DMU, and diminishing returns can be obtained by an additional linear constraint (Seiford and Thrall 1990).

Furthermore, the corresponding surface can be generated to form the lower convex hull of the functional values of the point of relevant pairs of decisions. The envelope function containing the relevant points is denoted by $Plf^E(z_1, z_2)$. Figure 1 illustrates a surface where the relevant decision pairs are located at their respective grid points.

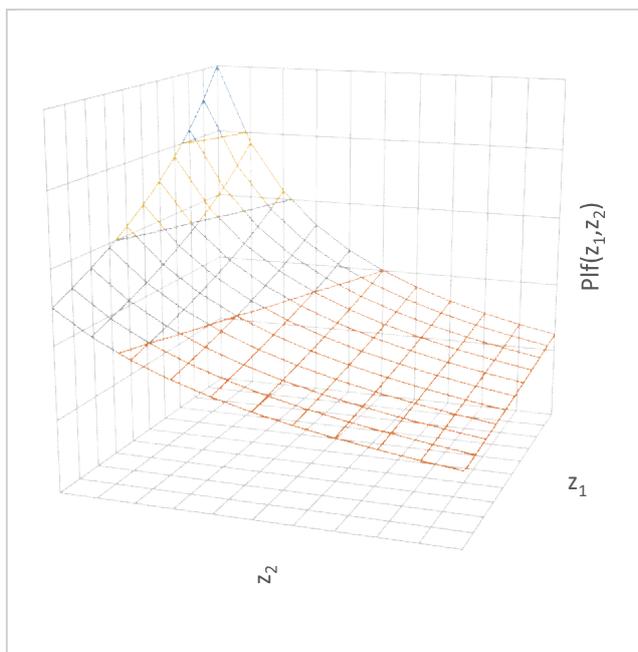

**Fig. 1**
Visualization of the probability of loss as a function of direct costs

*Proposition 1.* The total cost of any transaction has a minimum value.

*Proof:* In the domain of relevant points the transaction cost is defined as $Tcf(e, z_1, z_2) = z_1 + z_2 + Plf^E(z_1, z_2)e$. $Tc(e, 0,0) = Pl^E(0,0)e$ and direct costs can be arbitrarily high. The direct costs of the transaction are formed of increasing addends while the extent of loss decreases. The aggregated addends of the sum of direct costs of actors and losses are bounded from below.

The minimum transaction cost is located at the point of lowest $(z_1, z_2)$ values where the additional cost of the next possible choice of both actors is higher than the decrease in loss given by the probability of loss multiplied by the extent of exposure. There may also be more than one minimum point. A single minimum point might be located at $z_1 = 0, z_2 = 0$ if the cost of all choices of actors is higher than losses which may be saved. Whatever the shape of the loss ratio function, the exposure can be sufficiently low to render $z_1 = 0, z_2 = 0$ to be the minimum point of the transaction cost.

Williamson addresses the problem of optimization by saying 'Economizing on transaction costs essentially reduces to economizing on bounded rationality while simultaneously safeguarding the transactions in question against the hazards of opportunism. Holding governance structure constant, these two objectives are in tension, reduction in one commonly results in an increase in the other.' (Williamson 1979, pp. 245-246). Our model generalizes the trade-off between direct cost and hazard regardless if it is associated with opportunism or any other external factor.

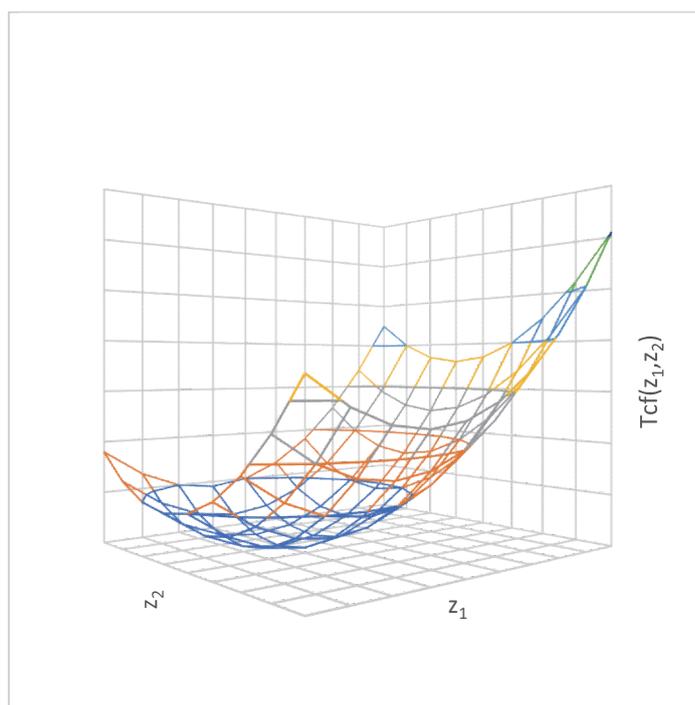

**Fig. 2**
Visualization of transaction costs as a function of choices of actors at a given exposure level

Optimal choice is also considered as a function of exposure and the model thus assumes that actors decide on the extent of exposure before initiating a transaction. Therefore, from the perspective of execution it is not considered as a decision variable.

The extent of motivation to execute a transaction is a prime external factor of the model which assumes, however, that even if the transaction cost is optimized, the portion to be paid by actors may emerge to be too high for one or for both so that it is subsequently rejected. A specific case of this is presented by Anderlini and Felli's (2006) game theory model of a 'hold-up' problem to indicate that *ex ante* transaction costs associated with negotiations may be higher for one or for both parties than the actual overall gain from the transaction.

## 3.3. Definition of the Transaction Cost Game

In this section, the execution of transaction as a two-player decision problem is examined in the context of a transaction cost game. Transaction costs are assumed to be shared by actors in that there is no external cost bearer. The common goal of both actors is therefore to execute the transaction with the least possible cost.

*Definition*: The *transaction cost sharing rule* assigns the proportion of total transaction cost to be paid by each actor for any possible combination of decisions for one given transaction cost type. Since exposure levels also affects the transaction cost the rule may vary due to its actual extent. The transaction cost sharing rule is thus defined as $Cs(Ttype, e)$. Elements of the rule comprise of distributions for all possible choice combinations $c_1(e, ch_i^1, ch_j^2)$, $c_2(e, ch_i^1, ch_j^2) \in Cs(Ttype, e)$     $i = 1, ..., n;$ $j = 1, ..., m$ thereby satisfying the following conditions.

$$0 \leq c_1(e, ch_i^1, ch_j^2) \leq 1, \quad 0 \leq c_2\left(e, ch_i^1, ch_j^2\right) \leq 1,$$

$$c_1\left(e, ch_i^1, ch_j^2\right) + c_2(e, ch_i^1, ch_j^2) = 1.$$

It is assumed that the Players know the transaction cost sharing rule.

The components of the transaction cost paid by Actors 1 and 2 are given as:

$$TcS_1\left(e, ch_i^1, ch_j^2\right) = c_1\left(e, ch_i^1, ch_j^2\right)Tc\left(e, ch_i^1, ch_j^2\right),$$

$$TcS_2\left(e, ch_i^1, ch_j^2\right) = c_2\left(e, ch_i^1, ch_j^2\right)Tc\left(e, ch_i^1, ch_j^2\right).$$

The normal form of the game proceeds as follows:

*Player 1* and *Player 2* are the Actors executing the transaction in the two-player game.

*Players' strategies* are formed of decisions made regarding the level of effort they make to reduce the probability of loss of the transaction: $ch_i^1$ and $ch_j^2$.

Pay-off functions are determined by applying the transaction cost sharing rule $Cs(Ttype, e)$ to the respective choices of players. Pay-offs in this game are negative, so we may use the term of pay-in functions instead, we prefer however to keep the traditional usage. If $(ch_i^1, ch_j^2) \notin Pch$ results the transaction is not executed, and the players pay the direct cost of their choices, so their pay-offs are $P_1 = -c_1(z_1(ch_i^1))$, and $P_2 = -c_2(z_2(ch_j^2))$, respectively.

If, however $(ch_i^1, ch_j^2) \in Pch$ then the transaction is executed, and players share the total transaction cost by applying the transaction cost sharing rule, so their respective payoffs are

$$P_1 = -c_1\left(e, ch_i^1, ch_j^2\right)Tc\left(e, ch_i^1, ch_j^2\right),$$

$$P_2 = -c_2\big(e, ch_i^1, ch_j^2\big)Tc\big(e, ch_i^1, ch_j^2\big).$$

The solution or equilibrium of the game is determined by the transaction cost sharing rule. Since the players have a finite number of strategies, an equilibrium exists with any given transaction cost sharing rule. The game however may not have a pure strategic equilibrium. The transaction cost game is a non-zero-sum game since the cost of the transaction depends on the strategies of the respective players. The following section examines how the transaction cost sharing rule influences the transaction cost.

### 3.4. Optimizer Transaction Cost Sharing Rule

*Definition*: A transaction cost sharing rule is *optimizer* if upon its application due to any decision of one of the players the best decision of the alternate player is to choose the conditional minimum point, or if there are additional minimum points, any of them.

If a transaction cost sharing rule is optimizer, the Nash-equilibrium of the transaction cost game is the minimum point or any of the minimum points of total transaction cost.

*Proposition.* The optimizer transaction cost sharing rule always exists for pure strategies whether decisions are sequential or simultaneous.

*Proof.* Since the transaction cost is higher at any point other than the conditional optimum point(s) the transaction cost sharing rule may be used to charge higher costs to the other player when his/her choice is not at the conditional minimum point.

*Remark.* As previously proven, pairs of combinations that are not located on the convex hull cannot be optimal at any given exposure level. Thus, transaction cost pertaining to these pairs are higher than the minimum level so the transaction cost sharing rule can be applied to charge higher costs to both players than their cost levels at the minimum point.

*Proposition.* Any distribution of the transaction cost related to the minimum point can be an element of an optimized transaction cost sharing rule.

*Proof.* Since the transaction cost is higher at any point other than the conditional optimum point(s) the transaction cost sharing rule may be used to charge higher costs to a player when his/her choice is not located at the minimum point regardless of the extent of the transaction cost to be paid by him/her at the equilibrium/minimum point.

Applying any fixed share $0 < c_1 < 1$ for every pair of combinations would result in an optimizer transaction cost sharing rule. A value of 0 or 1 would eliminate the motivation of one of the players to make any effort to reduce the probability of loss.

Examples of the notions defined to this point are given in the following. In the examples below there are only relevant and feasible pairs of choices therefore the actions taken by the players can be represented by their direct cost.

### Example 1: Optimum as a function of exposure

The $Plf^E(z_1, z_2)$ function of a transaction type is shown in Table 1. The cost of choices of both players are 0, 1 and 2 indicated on the vertical and horizontal axes. The values in the table consist of probabilities of loss attributed to the possible combinations of efforts by both players.

**Table 1**
Direct costs and values of a probability loss function

|  | | Player 2 | | |
|---|---|---|---|---|
| | $z_1$ | 0 | 1 | 2 |
| Player 1 $z_2$ | 0 | 1.00 | 0.05 | 0.04 |
| | 1 | 0.05 | 0.03 | 0.02 |
| | 2 | 0.04 | 0.02 | 0.01 |

Source: The authors' numeric example

Taking a transaction, the exposure of which is 60, the transaction cost of the combinations of choices is shown in Table 2. The value of each cell is formed of the sum of the direct costs of efforts of each player and the product of the probability of loss and exposure. For example, the value in column 2, row 3 is 1 + 2 + 0.02*60= 4.2. If both players do not make any effort to lower the probability of loss (0,0), the total exposure is lost. The total optimum is 3.8 at (1,1), therefore any other combination of choices would result in a higher transaction cost.

**Table 2**
Direct costs and total transaction cost of a transaction cost function

|  | | Player 2 | | |
|---|---|---|---|---|
| | $z_1$ | 0 | 1 | 2 |
| Player 1 $z_2$ | 0 | 60.0 | 4.0 | 4.4 |
| | 1 | 4.0 | **3.8** | 4.2 |
| | 2 | 4.4 | 4.2 | 4.6 |

Source: The authors' numeric example

In order to demonstrate how economies of scale have an impact at the optimum the same transaction cost type shown above is used. If exposure is equal to 1 the total optimum is located at (0,0), thereby indicating that even the smallest possible effort of both players exceeds the decrease in expected loss, see Table 3.

**Table 3**
Direct costs and total transaction cost of a transaction cost function

|  | | Player 2 | | |
|---|---|---|---|---|
| | $z_1$ | 0 | 1 | 2 |
| Player 1 $z_2$ | 0 | **1.0** | 1.1 | 2.0 |
| | 1 | 1.1 | 2.0 | 3.0 |
| | 2 | 2.0 | 3.0 | 4.0 |

Source: The authors' numeric example

If the exposure level is 120, the total optimum is located at (2,2) because higher expenditure levels are worthwhile in order to reduce the probability of loss, see Table 4.

**Table 4**
Direct costs and total transaction cost of a transaction cost function

|  | | Player 2 | | |
|---|---|---|---|---|
|  | $z_1$ | 0 | 1 | 2 |
| $z_2$ | | | | |
| Player 1 | 0 | 120.0 | 7.0 | 6.8 |
|  | 1 | 7.0 | 5.6 | 5.4 |
|  | 2 | 6.8 | 5.4 | **5.2** |

Source: The authors' numeric example

## Example 2: Transaction cost sharing rule and equilibrium

By taking the probability of loss function previously illustrated with a permitted exposure level of 60, the following simple transaction cost sharing rule associates the share of transaction cost attributed by Player 1 $c_1\left(e, ch_i^1, ch_j^2\right)$ to each combination of choices. Since $c_2(e, ch_i^1, ch_j^2) = 1 - c_1\left(e, ch_i^1, ch_j^2\right)$ Player 2's share is also given therefore in tables illustrating transaction cost sharing rules we only show the proportion to be paid by Player 1.

**Table 5**
Transaction cost sharing rule

|  | | Player 2 | | |
|---|---|---|---|---|
|  | $z_1$ | 0 | 1 | 2 |
| $z_2$ | | | | |
| Player 1 | 0 | 0.5 | 0.5 | 0.5 |
|  | 1 | 0.5 | 0.5 | 0.5 |
|  | 2 | 0.5 | 0.5 | 0.5 |

Source: The authors' numeric example

By applying the rule of Table 5, the pay-off functions for each Player are given in Table 6. Each cell indicates the respective pay-off levels for both Players, i.e. the transaction cost to be paid by each. The decision problem of the Players is represented by bimatrix games here and in the consecutive examples.

**Table 6**
Bimatrix pay-off function of a transaction cost game

|  |  | Player 2 | | | | | |
|---|---|---|---|---|---|---|---|
|  |  | 0 | | 1 | | 2 | |
|  | 0 | 30.0 | 30.0 | 2.0 | 2.0 | 2.2 | 2.2 |
| Player 1 | 1 | 2.0 | 2.0 | **1.9** | **1.9** | 2.1 | 2.1 |
|  | 2 | 2.2 | 2.2 | 2.1 | 2.1 | 2.3 | 2.3 |

Source: The authors' numeric example

The equilibrium is at (1,1), which is the total optimum level as well, thus the transaction cost sharing rule acts as an optimizer.

Application of the transaction cost sharing rule in Table 7 using the same probability of loss function and exposure level would result in the pay-off functions shown by Table 8.

**Table 7**
Transaction cost sharing rule

|  |  | Player 2 | | |
|---|---|---|---|---|
| $z_2$ \ $z_1$ |  | 0 | 1 | 2 |
|  | 0 | 0.5 | 0.5 | 0.9 |
| Player 1 | 1 | 0.1 | 0.3 | 0.9 |
|  | 2 | 0.3 | 0.1 | 0.5 |

Source: The authors' numeric example

**Table 8**
Bimatrix pay-off function of a transaction cost game

|  |  | Player 2 | | | | | |
|---|---|---|---|---|---|---|---|
|  |  | 0 | | 1 | | 2 | |
|  | 0 | 30.0 | 30.0 | 2.0 | 2.0 | 4.0 | 0.4 |
| Player 1 | 1 | 0.4 | 3.6 | 1.1 | 2.7 | 3.8 | 0.4 |
|  | 2 | 1.3 | 3.1 | 0.4 | 3.8 | **2.3** | **2.3** |

Source: The authors' numeric example

The equilibrium is at (2,2) while the total optimum is still at (1,1). The transaction cost part to be paid by player 1 is too high at (0,2) and (1,2), and the transaction cost part to be paid by player 2 is too high at (2,0) and (2,1) resulting in (2,2) being the equilibrium.

The following example (see Tables 9 and 10) illustrates a transaction cost sharing rule that, if applied at the same $Plf^E(z_1, z_2)$ function and exposure level, would result in no equilibrium in the subset of pure strategies. The game has an equilibrium in the set of mixed strategies, which cannot however be located at the total optimum point.

**Table 9**
Transaction cost sharing rule

|  | | Player 2 | | |
|---|---|---|---|---|
| | $z_1$ | 0 | 1 | 2 |
| $z_2$ | | | | |
| | 0 | 0.5 | 0.5 | 0.9 |
| Player 1 | 1 | 0.1 | 0.3 | 0.2 |
| | 2 | 0.3 | 0.1 | 0.5 |

Source: The authors' numeric example

**Table 10**
Bimatrix pay-off function of a transaction cost game

| | | Player 2 | | | | | |
|---|---|---|---|---|---|---|---|
| | | 0 | | 1 | | 2 | |
| | 0 | 30.0 | 30.0 | 2.0 | 2.0 | 4.0 | 0.4 |
| Player 1 | 1 | 0.4 | 3.6 | 1.1 | 2.7 | 0.8 | 3.4 |
| | 2 | 1.3 | 3.1 | 0.4 | 3.8 | 2.3 | 2.3 |

Source: The authors' numeric example

## 3.5. The magnitude and balance of regret of not making the optimal choice

Since the distribution of transaction cost at the optimum point is not defined by the optimizer criterion, it can be unequal in that the total cost might be allocated to any one of the two players.

If the distribution of transaction costs at the minimum point is unequal, the strength of incentive not to make a suboptimal choice can be weaker for the player paying the higher share since the additional cost allocated to him/her when departing the optimum position can be lower. Regret of departing from the optimum point can be maximized for both players in such a way that it is equally balanced. If this is not possible, the difference between the transaction costs paid by each party can be minimized. For the purpose of formal and simple presentation of this problem, exposure from formulas is excluded. The concrete solution however would vary according to the volume of exposure.

This requirement is met if the transaction cost sharing rule satisfies the following criteria:

$$\left| \left( c_1(ch^1_{i_n}, ch^2_{j_o}) Tc(ch^1_{i_n}, ch^2_{j_o}) - c_1(ch^1_{i_o}, ch^2_{j_o}) Tc(ch^1_{i_o}, ch^2_{j_o}) \right) - \left( c_2(ch^1_{i_o}, ch^2_{j_n}) Tc(ch^1_{i_o}, ch^2_{j_n}) - c_2(ch^1_{i_o}, ch^2_{j_o}) Tc(ch^1_{i_o}, ch^2_{j_o}) \right) \right| \to min$$

where:

- $\left( ch^1_{i_o}, ch^2_{j_o} \right)$ is the optimal pair of choices;
- $\left( ch^1_{i_n}, ch^2_{j_o} \right)$ is the pair of choices when Player 1 departs the optimum for the point where the increase in transaction cost is at its lowest point;

- $\left( ch_{i_o}^1, ch_{j_n}^2 \right)$ is the pair of choices when Player 2 departs the optimum for the point where the increase in transaction cost is at its lowest point.

Application of the preceding criteria will either render departure from the optimum equally costly for both players or alternately maximize the cost for the player who pays less to do so.

The regret maximizing criterion may however not always be realistically applied. The transaction cost sharing rule may be designed in accordance with a "pay for your mistake" principle whereby the additional transaction cost caused by the player departing the optimum point should be incurred by him/her alone. Magnitude of regret may be and often is perceived as being located between the "maximum regret", and "pay for your mistake" principles.

### Example 3: Applying the maximum regret criterion in a transaction cost sharing rule

Changing the transaction cost sharing rule applied in Example 1 in a way to fulfil the maximum regret criterion would leave equilibrium of the game at (1,1). However, the penalty for leaving the total optimum point is maximized for both players and the equilibrium is consequently more robust (see tables 11 and 12).

**Table 11**
Transaction cost sharing rule

|  |  | Player 2 | | |
|---|---|---|---|---|
|  | $z_1$ | 0 | 1 | 2 |
| $z_2$ |  |  |  |  |
| Player 1 | 0 | 0.5 | 1.0 | 0.5 |
|  | 1 | 0.0 | 0.5 | 0.0 |
|  | 2 | 0.5 | 1.0 | 0.5 |

Source: The authors' numeric example

**Table 12**
Bimatrix pay-off function of a transaction cost game

|  |  | Player 2 | | | | | |
|---|---|---|---|---|---|---|---|
|  |  | 0 | | 1 | | 2 | |
| Player 1 | 0 | 3.0 | 3.0 | 4.0 | 0.0 | 2.2 | 2.2 |
|  | 1 | 0.0 | 4.0 | 1.9 | 1.9 | 0.0 | 4.2 |
|  | 2 | 2.2 | 2.2 | 4.2 | 0.0 | 2.3 | 2.3 |

Source: The authors' numeric example

## 3.6. Disputes aiming at changing the distribution of transaction costs

The transaction cost sharing rule sets the distribution of transaction costs, but players may dispute the result and attempt to change it in their own favour.

*Remark*: Another reason for starting a dispute can be that the actual sharing of transaction cost differs from what the rule originally proposes. In this core model however cases when actors initiate disputes to correct mistakes in sharing transaction cost are not considered.

Such actions also have direct costs denoted as $v_i^1, v_j^2$ which may modify the share of cost distribution $s_1^d(v_i^1, v_j^2)$ called the dispute function. In the two-player model $s_2^d(v_i^1, v_j^2) = 1 - s_1^d(v_i^1, v_j^2)$.

The initial value of the share is set by the transaction cost sharing rule depending on choices to decrease the probability of loss:

$$s_1^d(0,0) = c_1(e, ch_i^1, ch_j^2), \qquad s_2^d(0,0) = c_2(e, ch_i^1, ch_j^2).$$

The $s_1^d$ and $s_2^d$ functions represent the capacity of players to advocate their self-interest in the transaction cost sharing process. Hence the functions encapsulate the cost efficiency of efforts to decrease the share paid by each player.

Similarly, to the definition of the $Plf$ and $Plf^E$ functions, the non-effective pairs of expenditure levels are excluded from the domain of the share function $s_1^d$. Therefore, each player should consider all possible means of share reduction at a minimum cost level. It is also assumed that marginal share reduction decreases in direct analogy to the $Plf$ function, where the possible extent of decrease of loss also decreases. Hence $s_1^d$ consistently decreases and assumes a convex form in the first variable, and consistently increases and assumes a concave form in the second variable as illustrated in Figure 3.

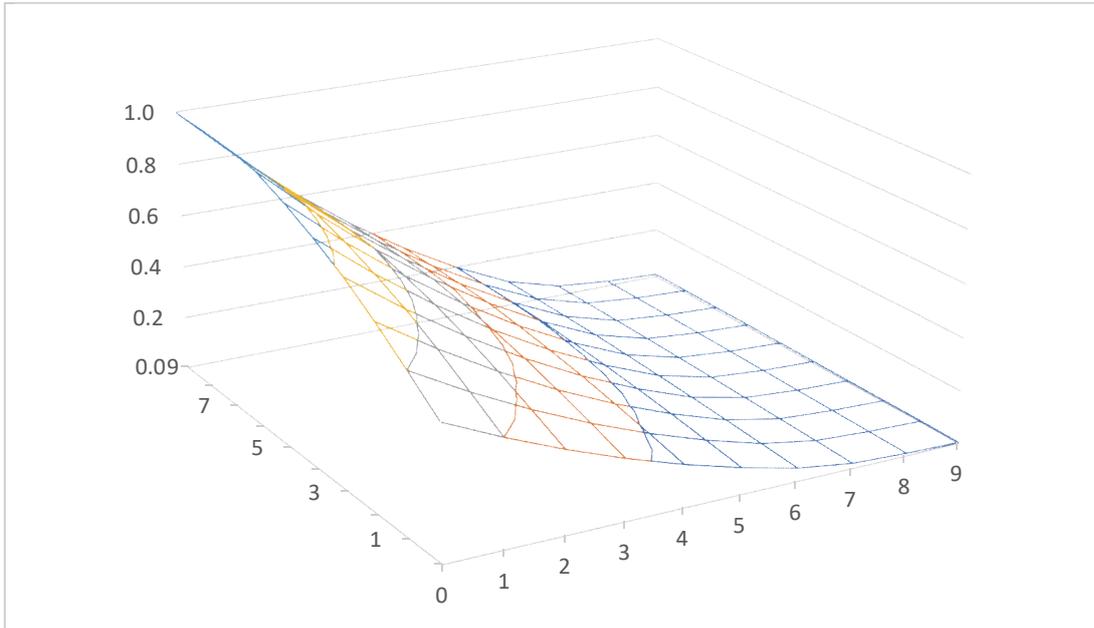

**Fig. 3**
Visualization of function $s_1^d(v_i^1, v_j^2)$

Each player may have to pay either more or less than the direct cost of their given choice. The proportions of the total direct cost of dispute to be paid by Player 1 and Player 2 are $d^1(v_i^1 + v_j^2)$ and $d^2(v_i^1 + v_j^2)$ respectively. $0 \leq d^1, d^2 \leq 1$ and $d^1 + d^2 = 1$.

The normal form of the game is outlined as follows:

The players decide the level of effort they make to change the distribution of transaction cost attributed to them by the transaction cost sharing rule. A choice of (0,0) would mean both accept the initial distribution.

*Strategies:*

*Player 1: $v_i^1$*

*Player 2: $v_j^2$*

*The pay-off functions are:*

*Player 1: $d^1(v_i^1 + v_j^2) + Tcs_1^d(v_i^1, v_j^2)$*

*Player 2: $d^2(v_i^1 + v_j^2) + Tcs_2^d(v_i^1, v_j^2)$*

## Example 4: Dispute about changing the proportion of transaction to paid by the Players

The values of one possible $s_1^d$ function are illustrated in the table below, correspondingly $s_2^d$=1-$s_1^d$. Each player can spend values of 0, 1 or 2 on decreasing the share of the transaction cost which must be paid. The transaction cost to be shared is assumed to have a value of 5, see Table 13.

**Table 13**
Direct costs and values of a dispute function

| Player 1 | $v_2$ \ $v_1$ | Player 2 | | |
|----------|-----|-----|-----|-----|
| | | 0 | 1 | 2 |
| | 0 | 0.5 | 0.8 | 0.9 |
| | 1 | 0.2 | 0.5 | 0.6 |
| | 2 | 0.1 | 0.4 | 0.5 |

Source: The authors' numeric example

Each Player pays the direct cost of respective efforts. The equilibrium of the game represented in the following table is at (1,1). Players would take efforts to change the transaction cost sharing process in their favour, but after expending equal efforts they pay the same proportion as they would have paid had they taken no action. Further stronger efforts to change the sharing process would increase the total cost paid by each player. The upper left 2×2 block of the pay-off matrix in Table 14 shows the same pattern as that of the prisoners' dilemma game, while economies of scale would not divert the equilibrium point form (1,1).

**Table 14**
Bimatrix pay-off functions of a dispute function

| | | Player 2 | | | | | |
|---|---|---|---|---|---|---|---|
| | | 0 | | 1 | | 2 | |
| | 0 | 2.5 | 2.5 | 4.0 | 2.0 | 4.5 | 2.5 |
| Player 1 | 1 | 2.0 | 4.0 | **3.5** | **3.5** | 4.0 | 4.0 |
| | 2 | 2.5 | 4.5 | 4.0 | 4.0 | 4.5 | 4.5 |

Source: The authors' numeric example

The transaction cost can be sufficiently low to make (0,0) the optimum and if the transaction cost sharing rule is optimizer, the equilibrium as well. If the same $s_1^d$ function were used where transaction cost to be shared is 3, players would not initiate a dispute to change their relative shares.

**Table 15**
Bimatrix pay-off functions of a dispute function

| | | Player 2 | | | | | |
|---|---|---|---|---|---|---|---|
| | | 0 | | 1 | | 2 | |
| | 0 | **1.5** | **1.5** | 2.4 | 1.6 | 2.7 | 2.3 |
| Player 1 | 1 | 1.6 | 2.4 | 2.5 | 2.5 | 2.8 | 3.2 |
| | 2 | 2.3 | 2.7 | 3.2 | 2.8 | 3.5 | 3.5 |

Source: The authors' numeric example

**Table 16**
Direct costs and values of a dispute function

| | | Player 2 | | |
|---|---|---|---|---|
| $v_1$ / $v_2$ | | 0 | 1 | 2 |
| | 0 | 0.8 | 0.9 | 0.95 |
| Player 1 | 1 | 0.3 | 0.5 | 0.6 |
| | 2 | 0.1 | 0.4 | 0.5 |

Source: The authors' numeric example

In this case the transaction cost to be shared is set at a value of 4, see Table 15.

In Table 16 the $s_1^d$ function is asymmetric. Player 1 would pay 80% of the transaction costs without making efforts to decrease his/her share. He/she can reduce his/her total cost by spending one extra unit and the equilibrium is (1,0) since player 2 would increase his/her cost by also spending one extra unit and reducing his/her part of the transaction cost by 0.8, see Table 17.

**Table 17**
Bimatrix pay-off functions of a dispute function

|  |  | Player 2 | | | | | |
|---|---|---|---|---|---|---|---|
|  |  | 0 | | 1 | | 2 | |
| Player 1 | 0 | 3.2 | 0.8 | 3.6 | 1.4 | 3.8 | 2.2 |
|  | 1 | **2.2** | **2.8** | 3.0 | 3.0 | 3.4 | 3.6 |
|  | 2 | 2.4 | 3.6 | 3.6 | 3.4 | 4.0 | 4.0 |

Source: The authors' numeric example

The following example illustrates that changing the rules also changes the decisions of players. The new rule is set so that if the dispute results in the same proportion as it was for the initial cost or is higher for the player who initiated the dispute, then he/she is liable to pay all dispute costs. This may be interpreted as such that if the player initiating the dispute lost the case, then he/she must take responsibility for additional costs. In this case, Players may make decisions sequentially, by Player 1 acting firstly and Player 2 secondly. The transaction cost to be shared is assumed to be 5.

**Table 18**
Direct costs and values of a dispute function

|  | $v_1$ | Player 2 | | |
|---|---|---|---|---|
| $v_2$ |  | 0 | 1 | 2 |
| Player 1 | 0 | 0.5 | - | - |
|  | 1 | 0.2 | 0.5 | 0.6 |
|  | 2 | 0.1 | 0.4 | 0.5 |

Source: The authors' numeric example

By examining the problem from the perspective of Player 1, he/she considers initiating the dispute, in certainty that he/she wins the case if the decision combinations are set at (1,0), (2,0) or (2,1). He/she realizes that (1,0) or (2,0) would be a better combination than (0,0) but also realizes that Player 2 would opt for (1,1) instead of (1,0), and (2,2) instead of (2,0). Therefore, the best choice for Player 1 is not to initiate a dispute and the equilibrium is at (0,0).

**Table 19**
Bimatrix pay-off functions of a dispute function

| | | Player 2 | | | | | |
|---|---|---|---|---|---|---|---|
| | | 0 | | 1 | | 2 | |
| | 0 | **2.5** | **2.5** | - | - | - | - |
| Player 1 | 1 | 1.0 | 5.0 | 4.5 | 2.5 | 6.0 | 2.0 |
| | 2 | 0.5 | 6.5 | 2.0 | 6.0 | 6.5 | 2.5 |

Source: The authors' numeric example

Taking into account the cost of dispute too, the grand total of total transaction cost in this model is:

$$Tc\left(e, ch_i^1, ch_j^2\right) = z_1(ch_i^1) + z_2(ch_j^2) + Pl\left(ch_i^1, ch_j^2\right)e + v_1 + v_2$$

If institutions discourage disputes which result in the same or worse distributions of transaction costs for the player initiating the dispute as in the initial case, and the capacity of players to represent their self-interest is evenly balanced, then players would not initiate cost sharing disputes.

## The transaction cost sharing rule, and the rules of dispute as regulators of actors' behaviour

As it was shown earlier the transaction cost sharing rule serves as the pay-off function of the transaction cost game. Different rules may imply different strategies for the Players. Rules may stimulate Players to choose the optimum point of the transaction, but also may drive them to make choices that increase the total transaction cost. The capacity to represent the self interest of the Players and the rules of dispute also have an impact on the total transaction cost. They may eliminate disputes and the cost of dispute or may make them a good choice increasing total transaction costs. Thus, the model shows the importance of rules or in other words institutions.

There are many notable issues not discussed in the framework of the core model presented in this paper which may include the following:

- The role of intermediary organizations in the form of actors such as banks, traders or lawyers in transactions is essential and indication of their functions in an extended model is indispensable. Similarly, it is important to investigate the impact of rent-seekers upon total transaction costs.
- A multiplayer version of the model would provide a more sophisticated picture of transactions and would lead to more precise conclusions on how efficient institutions should work in practice.
- Finally, the impact of transaction costs on allocation of resources should be formally modelled.

## 4. Conclusion

In terms of formulating means of decreasing transaction costs, decision makers should consider that losses are included as well as direct transaction costs. Moreover, since there is a trade-off between the two components, achieving lower transaction costs is a problem of optimization. Reducing direct transaction costs might increase losses by a higher volume than the decrease which results in higher total transaction costs. Following from the existence of optimum, there is a minimum level of transaction cost in any institutional environment therefore mitigating transaction costs and thus improving the efficiency of resource allocation can be achieved by the reform of institutions.

The optimal decision on executing a transaction may differ according to the extent of exposure. Thus, several acts of legislation make a distinction as to the volume of transactions. Importantly, the European Union PSD2 directive allows for exemption from application of security requirements of strict customer authentication, subject to conditions based on the level of risk and the amount and recurrence of the payment transaction. The lower the probability of loss of a transaction type as in for example the case of remote electronic card-based payments, the higher will be the transaction size below which simple authentication in the exemption threshold value can be applied.

In the model presented in this paper the level of uncertainty is not fixed. Instead, it can be reduced by additional effort. Uncertainty is quantified as the probability of loss incurred while executing the transaction.

Williamson (1979) suggests that markets efficiently organize occasional or recurrent transactions defined as non-specific. In this paper, by having introduced the concept of transaction type it is argued that both procedures and institutions may be subject to optimization. Hence regulation should aim at generating and maintaining the motivation of actors to make decisions necessary for executing transactions at minimum total costs.

Based on the analytical framework of the model, features of ideal institutional systems which may determine total transaction costs are identified. Transaction costs respectively depend on:

- How efficiently the probability of loss can be reduced by the efforts of actors/players.

The impact on actors' choices is determined by institutions in the form of rules and norms and by the technology of communication. In an economy characterized by secure property rights, efficient public administration and robust legal compliance the probability of loss can be reduced to a relatively low level with modest costs. Insecure property rights, inefficient government services and legal uncertainty may alternately cause higher losses even at higher direct cost levels.

- The efficiency of transaction cost sharing rules.

As previously demonstrated, transaction cost sharing rules can be defined as either an optimizer or as a non-optimizer. If the total optimum is the equilibrium of the transaction cost game the total costs of exchange are therefore lower, while non-optimizer rules will result in suboptimal decisions resulting in higher costs.

- The balance of incentives to choose the optimum for players.

Even if the transaction cost sharing rule is defined as an optimizer, one of the players might have to pay a very high share at the equilibrium point of the game and the price of departing it might be very low thus increasing the probability of wrong decisions. A sufficiently balanced level of regret of not choosing the total optimum level would result in relatively lower total transaction costs.

- The portion of shares of cancelled transactions.

A form of institutional uncertainty arises when the rules of executing a transaction are not sufficiently clear and the choices of players are not elements of the set of possible combinations. The direct costs of failed transactions are still paid by players themselves or other parties. Thus, the total cost of exchange in terms of market operation is higher than that of a lower rate of initiated but failed transactions.

- The cost of disputes aimed at changing the distribution of transaction costs.

Disputes increase total transaction costs whatever their outcome may be. In this model, neither of the players are expected to refrain from trying to represent their respective self-interests. However if the transaction cost sharing rule is set as a minimizer and regret of not choosing the optimum level is balanced, it is reasonable to minimize the cost of disputes. Three factors may contribute to this situation in the model presented here. Firstly, since a relatively high amount of transaction costs to be paid by a player would render starting a dispute the best choice for him/her, balanced distribution itself would reduce the number of disputes. Secondly, when the cost of the dispute is paid by the player who initiated it, if he/she cannot reduce his/her share of costs, fewer disputes would be initiated. Finally, if the capacity of players to represent their self-interest levels are evenly balanced, disputes are less likely to occur.

## 5. Closing remarks

The actors in this model possess imperfect information and one of the goals of their efforts is to improve their knowledge of factors determining the probability of loss. It is partly the decision of actors as to how well they are informed, and the level of information also has an optimum value.

The probability of emergence of loss function and transaction cost sharing rules are determined by the institutional and technological environments in which the transaction is executed. This includes the legal framework, informal rules and norms and the level of physical infrastructure and information technology. The model posits that regulators play a role in mitigating transaction costs given that even if only private sector actors are involved in executing a transaction, the institutional environment has an impact on total costs.

The model illustrates the underlying rationality behind what often seems to be market imperfection. The relationship between direct transaction costs and the probability of loss is one of the reasons for the phenomenon often referred to as bounded rationality.